\documentclass[a4paper,UKenglish]{lipics}
\usepackage{microtype}%if unwanted, comment out or use option "draft"
\usepackage{latexsym}
\usepackage{listings}
\usepackage{amsmath}
\usepackage{amsfonts}
\usepackage{verbatim}
\usepackage{bussproofs}
\usepackage{tikz-cd}
\usepackage{listings}

\newcommand{\El}{\textbf{El}}
\newcommand{\Prf}{\textbf{Prf}}
\newcommand{\G}{\Gamma}
\newcommand{\ts}{\vdash}
\newcommand{\Tworules}[6]
{\[ #3 \ \ \
    {{#1}\over{#2}}
    \ \ \ \ \ \ \ \ \ \
    #6 \ \ \
    {{#4}\over{#5}}
\]}

\newcommand{\Threerules}[9]
{\[ #3 \
    {{#1}\over{#2}}
    \ \ \ \ \
    #6 \
    {{#4}\over{#5}}
    \ \ \ \ \
    #9 \
    {{#7}\over{#8}}
\]}
\newcommand{\threerule}[6]              % special for LF-rules picture
{\[ {{#1}\over{#2}}
    \ \ \
    {{#3}\over{#4}}
    \ \ \
    {{#5}\over{#6}}
\]}

\newcommand{\C}{\mathcal{C}}
\newcommand{\E}{\mathcal{E}}
\newcommand{\U}{\mathcal{U}}
\newcommand{\Prop}{\textbf{Prop}}
\newcommand{\Type}{\textbf{Type}}
\newcommand{\ML}{Martin-L\"of}
\newcommand{\LTT}{$\text{LTT}_1$}

\newcommand{\N}{\textbf{N}}

\newcommand{\0}{\textbf{0}}

\title{Semi-simplicial Types in Logic-enriched Homotopy Type Theory}
\author[1]{Fedor Part}
\author[2]{Zhaohui Luo\thanks{Partially supported by research grants from Royal Academy of Engineering and the CAS/SAFEA International Partnership Program for Creative Research Teams.}}
\affil[1]{Royal Holloway, University of London\\
Email: fedor.part@gmail.com}
\affil[2]{Royal Holloway, University of London\\
Email: zhaohui.luo@hotmail.co.uk}
\authorrunning{F. Part and Z.Luo} %mandatory. First: Use abbreviated first/middle names. Second (only in severe cases): Use first author plus 'et. al.'

\Copyright{F. Part and Z.Luo}%mandatory, please use full first names. LIPIcs license is "CC-BY";  http://creativecommons.org/licenses/by/3.0/

% mandatory: Please choose ACM 1998 classifications from http://www.acm.org/about/class/ccs98-html . E.g., cite as "F.1.1 Models of Computation".
% mandatory: Please provide 1-5 keywords
\serieslogo{}%please provide filename (without suffix)
\volumeinfo%(easychair interface)
  {Billy Editor and Bill Editors}% editors
  {2}% number of editors: 1, 2, ....
  {Conference title on which this volume is based on}% event
  {1}% volume
  {1}% issue
  {1}% starting page number
\EventShortName{}
\DOI{10.4230/LIPIcs.xxx.yyy.p}% to be completed by the volume editor
\begin{document}

\maketitle

\begin{abstract}
The problem of defining Semi-Simplicial Types (SSTs) in Homotopy Type Theory (HoTT) has been recognized
as important during the Year of Univalent Foundations at the Institute of Advanced Study \cite{ias_page}. According to
the interpretation of HoTT in Quillen model categories \cite{DBLP:Awodey12}, SSTs are type-theoretic versions of Reedy fibrant
semi-simplicial objects in a model category and simplicial and semi-simplicial objects play a crucial role
in many constructions in homotopy theory and higher category theory. Attempts to define SSTs in HoTT lead
to some difficulties such as the need of infinitary assumptions which are beyond HoTT with only non-strict
equality types.

Voevodsky proposed a definition of SSTs in Homotopy Type System (HTS) \cite{hts_page}, an extension of HoTT with non-fibrant
types, including an extensional strict equality type. However, HTS doesn't have the desirable computational
properties such as decidability of type checking and strong normalization. In this paper, we study a
logic-enriched homotopy type theory, an alternative extension of HoTT with equational logic based on the idea
of logic-enriched type theories \cite{DBLP:AczelG06,DBLP:Luo06}. In contrast to Voevodsky’s HTS, all
types in our system are fibrant and it
can be implemented in existing proof assistants. We show how SSTs can be defined in our system and outline an
implementation in the proof assistant Plastic \cite{paul-luo:JAR00}.
\end{abstract}

\section{Introduction}
\label{sect:intro}

Homotopy Type Theory (HoTT) \cite{hottbook}, an extension of \ML's intensional Type Theory (MLTT), lies in the center of the research area that explores the striking connections between homotopy theory and type theory in the study of Univalent Foundations of mathematics. Providing a direct language for formalization of homotopy theory, HoTT inherits constructivity and some computational properties of MLTT so that proof assistants like Coq \cite{Coq:manualv8.1TYPES08} and Agda \cite{Agda:webTYPES08} can be used for proof verification and partial automatization. Sophisticated constructions from homotopy theory, whose complete formal description in set theory like ZF would be hopelessly cumbersome, can be expressed very concisely in HoTT, where the notion of a space is taken as primitive. For example, homotopy groups of spheres, fiber sequences, van Kampen theorem and many other things have been formalized in this way in HoTT \cite{hottbook}.

Unfortunately, these developments are obstructed by a substantial problem.  A type in HoTT is not only characterised by its elements, but by the whole structure of weak $\infty$-groupoid, generated by the type of paths. Correspondingly, a function in HoTT is not just one on elements of types, but a proper $\infty$-functor between $\infty$-groupoids and, as a result, the types form a weak $\infty$-category. However, $\infty$-functors and $\infty$-categories are meta-level notions
for HoTT, whereas attempts to define them internally lead to the need to encode an infinite
amount of coherence data, which is unclear how to do in HoTT. Only weak $n$-categories for concrete $n$ can be defined so far in HoTT. As a consequence, the notion of homotopy coherent diagram of types, which is a
$\infty$-functor from homotopy coherent nerve of a 1-category to the $\infty$-category of types, is also
problematic. Therefore, no general notion of homotopy limit can be formulated in HoTT, although for freely
generated diagrams it is possible \cite{AKL:HomLimCoq}.

A particular case of the problem is the internalisation of the notion of (semi-)simplicial objects
in a type universe $\U$ or, in short, (semi-)simplicial types. A simplicial type is the family
of types $X_i\colon \U$ ($i\in\omega$), together with maps $d_n^i:X_{n+1}\rightarrow X_n$
and $s^i_n:X_n\rightarrow X_{n+1}$, where $X_n$ should be thought of as the type of
$n$-dimensional simplices, $d_n^i$ assigns to a simplex $x:X_{n+1}$ the $n$-dimensional
simplex, which should be thought of as $i$-th face of $x$, and $s^i_n$ assigns to a simplex $x:X_n$
its degenerated version at the dimension $n+1$. Formally, $d_n^i$ and $s^i_n$ are just some maps that
satisfy certain equational conditions which, if expressed by means of the path-equality in HoTT, 
should be accompanied by coherence conditions for these paths and this cannot be expressed in HoTT.
%This would exactly define homotopy coherent diagrams of the form $\Delta^{op}\rightarrow \U$,
%where $\Delta$ is the simplicial category.
%WHY HERE?
Among many other things, simplicial types may serve as a useful tool for talking about weak $\infty$-categories
in HoTT in terms of complete Segal spaces \cite{rezk:css}. 
%Under
%certain conditions, simplicial types define complete Segal spaces, which are $\infty$-categories, where
%$X_0$ is the type of objects $X_1$ is the type of 1-morphisms, $X_2$ is the type of 2-morphisms and
%so on.
%Then, as soon as homotopy coherent nerve and universe of types is realised as a complete Segal Space,
%one may define the notion of homotopy coherent diagram.

If one omits degeneracy maps in the definition of simplicial types, one would obtain Semi-Simplicial
Types (SSTs). It simplifies the definition, but is still interesting as SSTs may be used, for example, to define
complete semi-Segal spaces, which are $\infty$-categories without identities. What seems to be missing in HoTT while trying to define SSTs inductively is some kind of proof-irrelevant or strict equality.
Two definitions of SSTs have been proposed independently by Herbelin \cite{Herbelin:SST} and Voevodsky \cite{VV:SST}. Both definitions rely on a notion of strict equality which, in the former case, is proof-irrelevant equality of the universe of propositions $\Prop$ in CIC\footnote{The Calculus of Inductive Constructions (CIC) is the type theory implemented in the Coq proof assistant. More accurately, the current Coq system \cite{Coq:manualv8.1TYPES08} implements the predicative CIC (pCIC) where Set has become a predicative universe (we omit the details here).} and, in the latter case, is the extensional equality of Voevodsky's Homotopy Type System (HTS) \cite{VV:SST}. Unfortunately, CIC with the proof-irrelevant equality in $\Prop$ is known to be inconsistent with univalent universes. HTS extends HoTT with auxiliary types, which do not carry the structure of weak $\infty$-groupoid and correspond to non-fibrant objects in Quillen model category. The major disadvantage of HTS is the lack of basic computational properties such as decidability
of type checking or strong normlization, which makes it not implementable in any existing proof
assistant.\footnote{Please also note that decidability of type checking is a basic requirement for a type theory to be a reasonable logical system and one may regard this as not only a desirable but necessary property.}

In this paper we exploit the ideas from Logic-enriched Type Theories (LTTs) \cite{DBLP:AczelG06,DBLP:Luo06} to study an extension of HoTT, HoTTEq, where all types are fibrant and SSTs are definable.\footnote{LTTs are related to the logic with dependent sorts \cite{Belo:DepLogic}, Cartmell's Generalized Algebraic Theories \cite{Cart:GAT}, Makkai's First Order Logic with Dependent Sorts \cite{Makkai:FOLDS} and Maietti's PhD thesis \cite{Maietti:thesis}. We omit a detailed analysis here.}  LTTs are type theories in which logical propositions and datatypes are completely separate. It is thus possible to introduce axioms or new rules of deduction without affecting the world of datatypes.  For instance, classical principles can be applied merely to logic without destroying the constructive nature of types (see, for example, \cite{DBLP:Luo06,AL:TOCL10}).  LTTs have been implemented in the Plastic proof assistant \cite{paul-luo:JAR00,plastic_page} where formalization tasks based on LTTs have been done including, for instance, formalization of Weyl's predicative foundation of mathematics \cite{AL:TOCL10}.

We shall consider a strict equality in an LTT-setting whose datatype part is HoTT.  Our system HoTTEq extends HoTT with two kinds of logical formulas, logical equality and universal quantification, and a new induction principle for $\N$, the type of natural numbers.  The inductive definition of SSTs in HoTTEq
may be related to Voevodsky's definition in HTS as follows. Let $T = \sum\limits_{x:\U}A(x)$, where $A(x)$ encodes some data over a type $x$. In HTS the type family $sst_n$ of all $n$-truncated SSTs is defined by recursion on natural numbers $sst\_rec\colon\N\rightarrow \sum\limits_{t:T}P(t)$, where $sst_n:=\pi_1(\pi_1(sst\_rec(n)))$ and $P(t)$
is essentially a predicate, which expresses the functoriality of some maps in $\pi_2(\pi_1(t))$ by means of the strict equality. As
$P(t)$ contains the strict equality, $\sum\limits_{t:T}P(t)$ is non-fibrant. In
HoTTEq such non-fibrant types are avoided by means of the new induction principle for natural numbers,
which allows one to define a function $sst\_rec\colon\N\rightarrow T$ and to prove that $\forall n:\N.P(sst\_rec(n))$
simultaneously.

\begin{comment}
We consider the system HoTTEq, which extends HoTT with two kinds of formulas: logical equality between
terms and universal quantification, and induction principles. Inductive definition of SSTs in HoTTEq
may be related to the one in HTS as follows. Denote $T:=\sum\limits_{x:\U}A(x)$, where $A(x)$ encodes
some data over a type $x$. In HTS the family of types $sst_n$ of all $n$-truncated SSTs
 is defined by a recursion on natural numbers $sst\_rec:\N\rightarrow \sum\limits_{t:T}P(t)$, where
$sst_n:=\pi_1(\pi_1(sst\_rec(n)))$ and $P(t)$
is essentially a predicate, which expreses functoriality of some maps in $\pi_2(\pi_1(t))$ by means of strict equality. As
$P(t)$ contains strict equality, $\sum\limits_{t:T}P(t)$ is non-fibrant. In
HoTTEq such non-fibrant types are avoided by means of induction principle for natural numbers,
which allows to define a function $sst\_rec:\N\rightarrow T$ and to prove a predicate $\forall n:\N.P(sst\_rec(n))$
on it simultaneously.
\end{comment}

The plan of the paper is as follows. Section 2 contains some background information on SSTs and LTTs. HoTTEq is described in Section 3. Section 4 demonstrates how SSTs can be defined in HoTTEq and Section 5 outlines an implementation in Plastic.

\section{Background: Semi-simplicial Types and Logic-enriched Type Theories}

Originally, \ML's intensional type theory (MLTT) was developed as a constructive foundation of mathematics
with meaning explanations based on the notions of canonical object and computation
\cite{ML:book84,ML:meaning85}.  Surprisingly, the system admits a non-trivial interpretation in abstract homotopy theory, with types as abstract spaces, terms as continuous functions and identity types as spaces of paths. This unveils intensional type theories as a general syntactic framework for formal reasoning about constructions of homotopy theory. Formally, such a theory is modelled in a category with the Quillen model structure, allowing to talk about homotopical constructions internal to the category \cite{DBLP:Awodey12}. A Quillen model structure consists of three classes of maps - fibrations, cofibrations and weak equivalences, which satisfy certain axioms \cite{QMC:book}. Examples of Quillen model categories
include categories of topological spaces, groupoids and simplicial sets.
%
%Relevant information about a space from the point of view of homotopy theory can be represented
%as a higher category. Namely, a space is the weak $\infty$-groupoid, where objects are points of
%the space, 1-morphisms are paths, 2-morphisms are homotopies between paths and so on.
%Similarly, there is a structure of weak $\infty$-groupoid on a type, which is internal to
%an intensional type theory and is generated by the \ML{} identity type as the \emph{type of paths} %\cite{BG:typesaregrp}.
%This property of intensional type theories makes them promising as formal systems for reasoning
%about higher categories.
%
%Reflexivity, symmetry and transitivity of \ML{} identity type correspond to
%constant path, inverse path and composition of paths.

In particular, HoTT extends MLTT to reflect some important properties of the simplicial set model \cite{VV:SSetModel} and, among other things, adds the Univalence Axiom, which is a type-theoretic version of the existence of object classifier in an elementary $\infty$-topos \cite{Lurie:book}.
HoTT hence provides a direct language for formalization of homotopy theory and higher category theory \cite{hottbook} but, as mentioned in the Introduction, there is an obstacle in defining (semi-)simplicial types, a notion to be introduced below.  We shall also review the LTT-framework whose ideas are used in this paper.

\subsection{Semi-simplicial types}
\label{sect:SST}

Let $\Delta$ denote the simplicial category with finite ordinals $Ob_{\Delta}=\{ [n]|n\in \N \}$ as objects and with morphisms generated by two classes of monotonic maps:
face maps  $d^i_n\colon [n]\hookrightarrow [n+1]$, omitting $i$ in the image, and degeneracy maps $s^i_n:[n+1]\rightarrow [n]$,
merging $i$ and $i+1$. A simplicial type is a simplicial object in a universe $\U$, that is a contravariant functor $X:\Delta^{op}\rightarrow \U$,
corresponding to a sequence of types $X([0]) :\U$, $X([1]):\U$, $X([2]):\U$, \dots, $X([n]):\U$, \dots,
together with maps $\bar{d}_n^i:X([n+1])\rightarrow X([n])$ and $\bar{s}_n^i:X([n])\rightarrow X([n+1])$, satisfying
equational conditions as determined by functoriality. $X([0])$ is the type of points, $X([1])$ the type of segments, \dots, and $X([n])$ the type of $n$-simplices, and so on. The maps $\bar{d}_n^i$ assign a face to $(n+1)$-simplices by omitting the $i$-th vertex and the maps $\bar{s}_n^i$ form the degenerated $(n+1)$-simplices out of $n$-simplices by repeating the $i$-th vertex.

%Defining ST in type theory is a first step towards type theoretic developments of a theory of $\infty$-categories
%in terms of Complete Segal Spaces ([some ref]) and a theory of homotopy limits in terms of simplicial objects ([some ref]).

%As a warm-up one may ask, whether Semi-Simplicial Types can be defined in HoTT.

In the case of semi-simplicial types (SSTs), one only needs to define the notion of a functor from the semi-simplicial category, which is a subcategory of the simplicial category on face maps. In this
paper we are mostly concerned with SSTs and shall use $\Delta$ to denote the semi-simplicial category, unless stated otherwise.

During the Year of Univalent Foundations at IAS \cite{ias_page} the following presentation of SSTs as a family of dependency structures of HoTT has been proposed.  Consider the following boundary map:
$$\bar{d}_n:=\bar{d}_n^0\times \dots \times \bar{d}_n^n:X([n+1])\rightarrow X([n])\times \dots \times X([n])$$
that assigns to each $(n+1)$-simplex a tuple of its $n$-faces. Then $X([n+1])$ is equivalent to the total space of fibration:
$$ X([n+1])\simeq \sum\limits_{x_0:X([n]),\dots,x_n:X([n])}(\bar{d}_n)^{-1}(x_0,\dots,x_n)$$
where fiber $(\bar{d}_n)^{-1}(x_0,\dots,x_n)$ is the type of all $(n+1)$-simplicies with boundary $(x_0,\dots,x_n)$ and
is non-empty iff $(x_0,\dots,x_n)$ forms a valid boundary. In the case that there's a type of all valid boundaries $bnd_n$,
$X([n+1])$ can be defined by means of the dependent type of fillings of boundaries $Y_n:bnd_n\rightarrow \U$:
$$X([n+1]):= \sum\limits_{x:bnd_n}Y_n(x)$$
Thus an SST is the following dependency structure $(Y_n)$:
\begin{equation} \label{eq:sst}
  \begin{split}
    Y_0\colon & \U \\
    Y_1\colon & Y_0 \rightarrow Y_0 \rightarrow \U \\
    Y_2\colon & \prod\limits_{a,b,c:Y_0}Y_1(a,b)\rightarrow Y_1(b,c)\rightarrow Y_1(a,c)\rightarrow \U \\
    \dots
  \end{split}
\end{equation}
In this way one may define $n$-truncated SSTs in HoTT for any \emph{concrete} $n$, but not for hypothised $n$. This only constitutes a meta-level
definition.

\subsection{Logic-enriched type theories}
\label{sec:LTT}

The concept of an LTT, an extension of the notion of type theory, was proposed by Aczel and Gambino in their study of type-theoretic interpretations of constructive set theory \cite{DBLP:AczelG06}. It provided them with flexibility to consider logics distinct from the propositions-as-types logic.
A type-theoretic framework, which formulates LTTs in a logical framework, has been proposed in
\cite{DBLP:Luo06} to support formal reasoning with different logical foundations.

An LTT is a dependent type theory such as MLTT extended with judgements of the following forms:
\begin{itemize}
  \item $\Gamma\vdash P\colon \Prop$, asserting that $P$ is a logical proposition under context $\Gamma$.
  \item $\Gamma\vdash p\colon \textbf{Prf}(P)$, asserting that $p$ is a proof of proposition $P$ (we often omit the $\textbf{Prf}$-operator.)\footnote{It is possible to consider a system without proof terms by only considering judgements of the form $\Gamma\vdash \phi_1,\dots,\phi_n \Rightarrow \phi\ true$, as in \cite{DBLP:AczelG06}.  Here we consider proof terms and they do not make any essential difference in most of the cases (and definitely not in this paper).}
\end{itemize}
In this paper, we shall employ the LTT-framework as studied in \cite{DBLP:Luo06} where a logical framework is used to specify LTTs, which is implemented in the proof assistant Plastic \cite{paul-luo:JAR00,plastic_page}.
A logical frmawork is a meta-language for specifying type theories, which is itself a dependent type theory. The rules of this framework for LTTs can be found in \cite{AL:TOCL10} (particularly its Appendix A).  Besides the kind \Type\ of types, we also have the kind \Prop\ whose objects are logical propositions.\footnote{Note that, different from Coq, \Prop\ is not a type and, hence, we do not automatedly have higher-order logic in LTTs.}  These kinds are governed by the following rules:
\Threerules
{\G\ valid}
{\G\ts \Type\ kind}
{}
{\G\ts A\colon \Type}
{\G\ts \El(A)\ kind}
{}
{\G\ts A=B\colon \Type}
{\G\ts \El(A)=\El(B)}
{}
\Threerules
{\G\ valid}
{\G\ts \Prop\ kind}
{}
{\G\ts P\colon \Prop}
{\G\ts \Prf(P)\ kind}
{}
{\G\ts P=Q\colon \Prop}
{\G\ts \Prf(P)=\Prf(Q)}
{}
The rules for valid context formation is as usual:
\Tworules
{}
{\langle\rangle\ valid}
{}
{\G\ts K\ kind\ \ \ x\not\in FV(\G)}
{\G,\ x:K\ valid}
{}
Please note, however, since there are new kinds \Prop\ and $\Prf(P)$, one can assume propositions and their proofs as well as types and their objects.  
There
are also parametric kinds of the form $(x:K)K'$, which are $\Pi$-constructors at the level of kinds.  They are also called $\Pi$-kinds, not to be confused
with $\Pi$-types: the former is meta-level constructs while the latter is at the object level. $\Pi$-types can be introduced as
follows:
\begin{align*}
	\Pi &\colon (A:\textbf{Type},B:(x:A)\textbf{Type})\textbf{Type} \\
	\lambda &\colon (A:\textbf{Type},B:(x:A)\textbf{Type},f:(x:A)B[x])\Pi[A,B] \\
	\mathcal{E}_{\Pi} &\colon (A:\textbf{Type},B:(x:A)\textbf{Type},C:(F:\Pi[A,B])\textbf{Type}, \\
	& f:(g:(x:A)B[x])C[\lambda[A,B,g]],z:\Pi[A,B])C[z]
\end{align*}
plus a computation rule which we omit here.

\begin{comment}
Before LTTs, dependently sorted systems have been studied in works of Cartmell on Generalized Algebraic
Theories ([]) and Makkai on First-Order Logic with Dependent Sorts ([]). Although both GAT and FOLDS were
inspired by MLTT, neither of them uses the notion of sort, directly corresponding to the notion of type of
dependent type theory. General notion of logic with dependent sorts, which includes GAT, FOLDS and LTT,
has been developed by P. Aczel and J. Belo ([]).
\end{comment}

%LTTs are related to the logic with dependent sorts \cite{Belo:DepLogic},
%which include Cartmell's Generalized Algebraic
%Theories \cite{Cart:GAT} and Makkai's First-Order Logic with Dependent Sorts \cite{Makkai:FOLDS}.

The nice property of LTTs is that the introduction of new logical axioms or logical rules doesn't affect world of types.  This has given birth to an interesting direction of research.  For instance, traditionally, the propositions-as-types logic in dependent type theories cannot be made classical
without introducing unwanted closed terms in the data-types such as that of natural numbers, which do not compute to any canonical objects. For example, in MLTT the law of Exluded Middle would generate an element of $\N+ (\N\rightarrow \0)$, which doesn't compute to any canonical element of the type.  As LTTs are free of such a problem, they are better suited for classical reasoning with type theory.  Work on the subject includes \cite{DBLP:Luo06,AL:TOCL10} and the Plastic proof assistant has been developed to support LTTs \cite{paul-luo:JAR00,plastic_page} and it was used to give computer checked
formalization of Weyl's classical predicative mathematics \cite{AL:TOCL10}.

\begin{comment}
In \cite{DBLP:Luo06} LTTs have been formulated in terms of the logical framework, which is implemented in Plastic.
A logical frmawork is a meta-language for specifying type theories, which is itself a dependent type theory. It deals with
\emph{kinds} such as the kind of all types \Type{} or kind of elements $El(A)$ of a type $A\colon \Type$. There
are also parametric kinds of the form $(\Delta)T$, where $\Delta$ is non-empty context of the form
$x_1\colon K_1,\dots, x_n\colon K_n$ and $K_1,\dots,K_n$ are kinds and $T$ is a type. The logical propositions of an LTT are objects of kind \Prop\
in this logical framework and, for each proposition $P\colon \Prop$, there is a kind $Prf(P)$ of its proofs. Contexts of LTTs may contain both object variables and proof variables.
As an example, in \cite{DBLP:Luo06} the system \LTT\, which is MLTT eriched with classical first-order logic, has been defined.
\end{comment}

In this work we investigate another interesting application of the LTT-framework: resolving the coherence problem in HoTT.
For this purpose we slightly diverge from what has been understood as an LTT in \cite{DBLP:AczelG06} or \cite{DBLP:Luo06} in the following sense.
\begin{itemize}
    \item We form HoTTEq by adding a logical equality and universal quantification, but it doesn't have full first-order logic of LTTs.  In another words, in HoTTEq, only equational reasining is considered and, in a sense, it is closely related to generalized algebraic theories.
    \item Logically equal terms in HoTTEq are interchangeble in both propositions and types, whereas
      in LTTs they are interchangeble only in propositions.
      %Equality of HoTT+LEq+Ind is intensional for
      %computational reasons. We also consider the system HoTT+ExtLEq+Ind, which has extensional equality,
      %as internal language.
    \item Induction principles of HoTTEq are more sophisticated than those in LTTs. In LTTs an inductive type
      is characterised by an elimination principle and logical axioms of induction. These are two
      separate mechanisms for constructing functions and proving propositions. In HoTTEq these
      mechanisms are merged into one.
\end{itemize}
However, having said the above, the key idea of LTTs that propositions and datatypes are separated plays a key role in the work to be described below.

\section{HoTTEq: HoTT Enriched with Logical Equality}

As in the LTT-framework (see \S\ref{sec:LTT}), HoTTEq extends HoTT with two forms of judgements: $\Gamma\vdash P\colon\Prop$ and $\Gamma\vdash p\colon P$.  Contexts of HoTTEq can contain both type/object variables and proposition/proof variables. We shall follow the rules for \LTT\ \cite{DBLP:Luo06}. In \LTT,
however, every context can be split: $\Gamma$ is equivalent to $(\Gamma_{\Type},\Gamma_{\Prop})$, where $\Gamma_{\Type}$ contains
type/object variables and $\Gamma_{\Prop}$ contains proposition/proof variables. This is because that, in traditional LTTs, propositions or their proofs do not occur in 
types or their objects.  For a type theory like \LTT, we have: $\Gamma_{\Type},\Gamma_{\Prop}\vdash J_{\Type}$ if, and only if, $\Gamma_{\Type}\vdash J_{\Type}$, where
$J_{\Type}$ is either $A:\Type$ or $a:A$. In contrast, in HoTTEq one can construct more objects with assumptions of the logical equality
between terms. This property will be crucial for the definition of SSTs.

We shall also have the rule for proof irrelevance:
\begin{prooftree}
	\centering
        \def\labelSpacing{12pt}
	\AxiomC{$\G\ts P \colon \Prop$}
	\AxiomC{$\G\ts p\colon P$}
	\AxiomC{$\G\ts q\colon P$}
        \RightLabel{(PI)}
	\TrinaryInfC{$\G\ts p\equiv q\colon P$}
\end{prooftree}
which states that any two proofs of a proposition are definitionally equal.

\subsection{Logical operators}

HoTTEq contains only two logical operators: equality $x =_A y\colon \Prop$ and universal quantification
$\forall x:\sigma.P(x)\colon \Prop$, where $\sigma$ is either a type or a proposition. The rules for the logical equality
are reminiscent of those of the \ML{} identity, where the $J$-like eliminator is duplicated for propositions and types.
Additionally, the logical equality is defined to satisfy function extensionality by the rule (LEqFE). The rules are as
follows:
\begin{prooftree}
	\centering
        \def\labelSpacing{12pt}
	\AxiomC{$\Gamma\vdash A \colon \Type$}
	\AxiomC{$\Gamma\vdash a\colon A$}
	\AxiomC{$\Gamma\vdash b\colon A$}
        \RightLabel{(LEqForm)}
	\TrinaryInfC{$\Gamma\vdash a =_A b\colon \Prop$}
\end{prooftree}

\begin{prooftree}
	\centering
        \def\labelSpacing{12pt}
        \AxiomC{$\Gamma\vdash A \colon \Type$}
        \AxiomC{$\Gamma\vdash a \colon A$}
        \RightLabel{(LEqIntro)}
        \BinaryInfC{$\Gamma\vdash \textbf{r}^A_a\colon a =_A a$}
\end{prooftree}

\begin{prooftree}
	\centering
        \def\labelSpacing{12pt}
	\AxiomC{$\Gamma, x:A,y:A,p:x =_A y \vdash T[x,y,p]\colon \Type$}
	\AxiomC{$\Gamma, x:A \vdash t(x):T[x,x,\textbf{r}]$}
        \RightLabel{\scriptsize(LEqElimT)}
	\BinaryInfC{$\Gamma, x:A,y:A,p:x =_A y \vdash \E^T_=(x,y,p,t)\colon T[x,y,p]$}
\end{prooftree}

\begin{prooftree}
	\centering
        \def\labelSpacing{12pt}
	\AxiomC{$\Gamma, x:A,y:A,p:x =_A y \vdash P[x,y,p]\colon \Prop$}
	\AxiomC{$\Gamma, x:A \vdash q(x):P[x,x,\textbf{r}]$}
        \RightLabel{\scriptsize(LEqElimP)}
	\BinaryInfC{$\Gamma, x:A,y:A,p:x =_A y \vdash \E^P_=(x,y,p,q)\colon P[x,y,p]$}
\end{prooftree}

\begin{prooftree}
	\centering
        \def\labelSpacing{12pt}
	\AxiomC{$\Gamma, x:A,y:A,p:x =_A y \vdash T[x,y,p]\colon \Type$}
	\AxiomC{$\Gamma, x:A \vdash t(x):T[x,x,\textbf{r}]$}
        \RightLabel{\scriptsize(LEqComp)}
	\BinaryInfC{$\Gamma, x:A \vdash Cmpt(x) \colon \E^T_=(x,x,\textbf{r},t)=_{T[x,x,\textbf{r}]}t(x)$}
\end{prooftree}

\begin{prooftree}
	\centering
        \def\labelSpacing{12pt}
	\AxiomC{$\Gamma\vdash  A \colon \Type$}
	\AxiomC{$\Gamma, x:A \vdash B(x) \colon \Type$}
        \RightLabel{\scriptsize(LEqFE)}
	\BinaryInfC{$\Gamma, f,g : F, p:(\forall x:A.f(x)=_{B(x)}g(x)) \vdash FE(f,g,p) \colon f=_F g$}
\end{prooftree}
where $F\equiv \prod\limits_{x:A}B(x)$.

And for the universal quantification (where $\sigma$ is either $\Type$ or $\Prop$):

\begin{prooftree}
	\centering
        \def\labelSpacing{12pt}
	\AxiomC{$\Gamma\vdash X \colon \sigma$}
	\AxiomC{$\Gamma, x:X \vdash P(x) \colon \Prop$}
        \RightLabel{(FAForm)}
	\BinaryInfC{$\Gamma\vdash \forall x:X.P(x)\colon \Prop$}
\end{prooftree}

\begin{prooftree}
	\centering
        \def\labelSpacing{12pt}
        \AxiomC{$\Gamma\vdash X\colon \sigma$}
        \AxiomC{$\Gamma, x:X \vdash p(x) \colon P(x)$}
        \RightLabel{(FAIntro)}
        \BinaryInfC{$\Gamma\vdash I_{\forall}(p)\colon (\forall x:X.P(x))$}
\end{prooftree}

\begin{prooftree}
	\centering
        \def\labelSpacing{12pt}
        \AxiomC{$\Gamma\vdash X\colon \sigma$}
	\AxiomC{$\Gamma\vdash p\colon (\forall x:X.P(x))$}
        \RightLabel{\scriptsize(FAElim)}
	\BinaryInfC{$\Gamma, x:X \vdash \E_{\forall}(p, x)\colon P(x)$}
\end{prooftree}

Using the eliminators $\E_=^P$ and $\E_=^T$ and the universal quantification, one can construct the transitivity proof: for $p\colon a =_A b$ and $q\colon b =_A c$, the proof 
$q\cdot p\colon a =_A c$.  Also, for $p\colon a =_A b$ and $(x:A)Y(x)\colon \sigma$, one can construct the substitution $subst^Y_p(y)\colon Y(b)$, where $y:Y(a)$.

The following straightforward lemmas assure that logical equality, defined in this way, behaves as expected. They will be required in Section 4.
\begin{lemma}
  For any type $A\colon \Type$, any type family $T\colon (A)\Type$ over A, if $p\colon x =_A y$, $q\colon y =_A z$, then $subst^T_q\circ subst^T_p=subst^T_{q\cdot p}$.
\end{lemma}
\begin{lemma}
  For any family of functional types $T\equiv(x:A)\prod\limits_{y:B}C(x,y)$, any object $a\colon A$, any function $f:\prod\limits_{y:B}C(a,y)$ and any equality proof 
  $p\colon a=_A b$, the following holds: $$\forall y:B. subst^T_p(f)(y)=subst^{C(-,y)}_p(f(y)).$$
\end{lemma}

\subsection{Induction}
\label{sec:ind}

In LTTs, an inductive type is characterised by an elimination rule that specifies how one can construct elements of other types out of the objects of the inductive type, and an induction rule that specifies  how propositions about objects of the inductive type can be proven.
For example, for the type of natural numbers \N{}, it does not only have the familiar elimination rule but have the following induction rule:
\begin{prooftree}
	\centering
        \AxiomC{$\Gamma,n:\N\vdash P_n\colon \Prop$}	
	\AxiomC{$\Gamma \vdash b \colon P_0$}
        \AxiomC{$\Gamma,n:\N, p:P_n\vdash ih \colon P_{n+1}$}
	\TrinaryInfC{$\Gamma, n\colon N\vdash Ind_{\N}(b,ih, n)\colon P_n$}
\end{prooftree}
This is fine as long as proof terms do not occur in types or their objects. In HoTTEq this property doesn't hold and the above form of induction becomes insufficient. This happens because, if inductive hypothesis
has the form:
$$n\colon N,t\colon T_n,p\colon P_n(t)\vdash ih_T(t, p)\colon T_{n+1}$$
$$n\colon N,t\colon T_n,p\colon P_n(t)\vdash ih_P(t, p)\colon P_{n+1}(ih_T(t,p))$$
where $T_n$ are types and $P_n(t)$ are propositions, then the elimination rule for $\N$ cannot be applied to
obtain a function of type $\prod\limits_{n:\N}T_n$. However note, that if propositions were types, this
inductive hypothesis could have been rewritten in terms of type family $\sum\limits_{t:T_n}P_n(t)$.
Thus, instead, we introduce a mechanism that not only constructs a function from an inductive type but, simulteniously, proves a property of the function (the premises are the same):
\begin{prooftree}
	\centering
	\AxiomC{$\Gamma \vdash b \colon T_0$}
        \AxiomC{$\Gamma \vdash pb \colon P_0(b)$}
        \noLine
	\BinaryInfC{$\Gamma, n\colon N,t\colon T_n,p\colon P_n(t)\vdash ih_T(t, p)\colon T_{n+1}$}
        \noLine
        \UnaryInfC{$\Gamma, n\colon N,t\colon T_n,p\colon P_n(t)\vdash ih_P(t, p)\colon P_{n+1}(ih_T(t,p))$}
	\UnaryInfC{$\Gamma, n\colon N\vdash \E_N^T(b,pb,ih_T,ih_P,n)\colon T_n$}
\end{prooftree}
and 
\begin{prooftree}
	\centering
        \AxiomC{$\Gamma \vdash b \colon T_0$}
        \AxiomC{$\Gamma \vdash pb \colon P_0(b)$}
        \noLine
	\BinaryInfC{$\Gamma, n\colon N,t\colon T_n,p\colon P_n(t)\vdash ih_T(t, p)\colon T_{n+1}$}
        \noLine
        \UnaryInfC{$\Gamma, n\colon N,t\colon T_n,p\colon P_n(t)\vdash ih_P(t, p)\colon P_{n+1}(ih_T(t,p))$}	
	\UnaryInfC{$\Gamma, n\colon N\vdash \E_N^P(b,pb,ih_T,ih_P,n)\colon P_n(\E_N^T(b,pb,ih_T,ih_P,n))$}
\end{prooftree}
with the following computation rules:
\begin{equation*}
  \begin{split}
    \Gamma \vdash & \E_N(b,pb,ih_T,ih_P,0)\equiv b
  \end{split}
\end{equation*}
\begin{equation*}
  \begin{split}
    \Gamma, n\colon N\vdash & \E_N(b,pb,ih_T,ih_P,n+1)\equiv 
    ih_T(\E_N^T(b,pb,ih_T,ih_P,n),\E_N^P(b,pb,ih_T,ih_P,n))
  \end{split}
\end{equation*}

%HoTT+LEq+Ind extends HoTT+LEq with these rules of induction. Note, that if propositions were types, these
%rules would have been derivable from elimination rule by applying it to the type family $\sum\limits_{t:T_n}P_n(t)$.

\section{Inductive Definition of Semi-Simplicial Types}
\label{sec:SST}

We shall first give an outline on how to define SSTs in HoTTEq and then describe the constructions in more details.

\subsection{Outline}

In this subsection we show how the inductive construction, as described by Voevodsky \cite{VV:SST}, of the tower of
dependency structures of SSTs (see (\ref{eq:sst}) at the end of \S\ref{sec:SST}) fails in HoTT, but can be done in HoTTEq.

The inductive procedure constructs $(n+1)$-truncated SSTs out of an $n$-truncated ones. Explicitly, types $sst_n$ of all 
$n$-truncated SSTs are defined inductively by assigning a type of $(n+1)$-simplices to every valid $n$-dimensional bounary of
a $(n+1)$-simplex:

$$sst_{n+1}\equiv \sum\limits_{x:sst_n}bnd_n^{n+1}(x)\rightarrow \U$$
where $bnd_n^m(x)$ is the type of $n$-boundaries of a $m$-simplex in $x$. The type $bnd_n^m(x)$ is defined inductively by
representing $bnd_{n+1}^m(x)$ as the type of pairs $(y,aug)$, where $y\colon bnd_n^m(x)$ and 
$aug\colon \prod\limits_{f:[n+1]\rightarrow [m]}Fill(rest_n^{f,x}(y))$ is an augmentation of $n$-boundary $y$, obtained by choosing
a $(n+1)$-simplex for every restriction of $y$ to $(n+1)$-dimensional subsimplex $f$ of a $m$-dimensional simplex. Here restriction
maps $rest_n^{f,x}(y)$ can only be defined inductively in the assumption of functoriality of them. Then functoriality can be defined
inductively only with further coherence assumptions and so on.

We avoid this obstacle in HoTTEq by using the induction principle for natural numbers as given in \S\ref{sec:ind}, which makes it possible
to prove functoriality mutually with the type construction. In \S\ref{sect:intro} $P$ denotes this functoriality predicate.
%Note that the family $sst_n$ of types of all $n$-\emph{truncated} SSTs
%for each $n$ is constructed in this way. The type of all SSTs can be defined as follows:
%$$sst := sst_0\times \prod\limits_{n:\N}sst_n\rightarrow sst_{n+1}.$$

%The type of all SST, including those containing simplices of arbitrary dimensions, can be
%constructed out of $sst_n$ by evident coinduction.

The rest of the section is devoted to the detailed exposition of how to construct the type family $sst_n$.

\subsection{Face maps}

We define combinatorics of semi-simplicial category in type theory by the type $\Delta(i,j)\colon \U_0$ of
all increasing functions between standard intervals $Stn(i)$ and $Stn(j)$, namely:

$$\Delta(i,j):= \sum\limits_{f:Stn(i)\rightarrow Stn(j)}is\_incr_T(f)$$
where:

$$Stn(i):= \sum\limits_{n:N}leq(n, i)$$
$$is\_incr_T(f):= \prod\limits_{n,m:Stn(i)}ls(n,m)\rightarrow ls(f(n),f(m))$$

Composition $-\circ_{\Delta} -$ of two face maps $(f,p_f)\colon \Delta(i,j)$ and $(g,p_g)\colon \Delta(j,k)$ is
defined as follows:

\begin{equation*}
  \begin{split}
    (g,p_g)& \circ_{\Delta}(f,p_f):= \\
     & (g\circ f, \lambda n,m:Stn(i).p_g(f(n),f(m))\circ p_f(n,m))
  \end{split}
\end{equation*}

Note that, by the $\eta$-rule, the associativity of composition holds \emph{definitionally}.

\subsection{Inductive construction}

We use induction for natural numbers HoTTEq to define simultaneously the data and a proof of a predicate
on this data. Data construction is as follows:
\begin{itemize}
    \item Type family $sst_n\colon \U_1$ of all $n$-truncated SST, such that $sst_{n+1}$ computes to
      $\sum\limits_{x:sst_n}bnd_n^{n+1}(x)\rightarrow \U_0$ and $sst_0 \equiv \U_0$.
    \item Type family $bnd_n^m(x:sst_n)\colon \U_1$ of $n$-boundaries of $m$-simplex in $x$, such that:

      \begin{itemize}
          \item $bnd_0^m(x:sst_0)\equiv Stn(m)\rightarrow x$.
          \item A $(n+1)$-boundary is a pair $(y,aug)$, consisting of $n$-boundary $y$ and it's augmentation
            $aug$. Precisely, $bnd_{n+1}^m(x, Fill)$, where
            $x:sst_n$, $Fill:bnd_n^{n+1}(x)\rightarrow \U_0$, computes to:
            $$\sum\limits_{y:bnd_n^m(x)}\prod\limits_{f:[n+1]\rightarrow [m]}Fill(rest_n^{f,x}(y))$$
      \end{itemize}

    \item Family of maps $rest_n^{f,x:sst_n}(y)\colon bnd_n^k(x)$, where $y:bnd_n^m(x)$, restricting $n$-boundary of
      $m$-simplex $y$ to the $n$-subboundary of face $f:\Delta(k, m)$ of $m$-simplex, such that:

      \begin{itemize}
        \item $rest_0^{f,x:sst_0}(y) \equiv y\circ \pi_1(f)$  %y:Stn(m)\rightarrow x)=_{\beta} y\circ \pi_1(f)$.
        \item Inductive step here deserves a comment as this is the point, where functoriality is required. At this stage
          the value of the following expression should be specified, where $(y,aug)$ is unfolded presentation of
          $(n+1)$-boundary $bnd_{n+1}^m(x)$:

          \begin{equation*}
            \begin{split}
	    rest_{n+1}^{f:\Delta(k,m),x}(y,aug)  %& y:bnd_n^m(x),\\
%            & aug \colon \prod\limits_{g:\Delta(n+1,m)}Fill(rest_n^{g,x}(y))
            \end{split}
          \end{equation*}

          This can be done by taking $n$-boundary $rest_n^{f,x}(y)\colon bnd_n^k(x)$ together with some augmentation of it, which
          should be of type $\prod\limits_{h:\Delta(n+1,k)}Fill(rest_n^{h,x}(rest_n^{f,x}(y)))$. But if we try to use given augmentation
          $\lambda h.aug(f\circ h)$, we would obtain a term of type $\prod\limits_{h:\Delta(n+1,k)}Fill(rest_n^{f\circ h,x}(y))$.
          Thus we require, that $rest_n$ satisfy the following functoriality condition:

%          $$func_n \colon (rest_n^{f\circ h,x} = rest_n^{h,x}\circ rest_n^{f,x})$$
\begin{equation*}
     \begin{split}
            func_n\colon \forall k, l, m:\N.\forall f:\Delta(k,l),& g:\Delta(l,m). \\
&\forall y:bnd_n^m(x).rest_n^{g\circ f,x}(y) = rest_n^{f,x}(rest_n^{g,x}(y))
     \end{split}
\end{equation*}
          The correct augmentation can be obtained by using substitution:
          $$\lambda h.subst^{Fill}_{func^{h,f}_n(y)}(aug(f\circ h))$$

      \end{itemize}
\end{itemize}

To complete the inductive definition, it remains to prove functoriality for $rest_0$ and for $rest_{n+1}$, provided it holds
for $rest_n$. Assume that $f:\Delta(k, l)$ and $g:\Delta(l,m)$. The base case is straightforward:
\begin{equation*}
     \begin{split}
       %y:Stn(m)\rightarrow x
       rest_0^{g\circ f,x:sst_0}(y) \equiv y\circ \pi_1(g\circ f) \equiv y\circ (\pi_1(g)\circ \pi_1(f))\equiv \\
\equiv (y\circ \pi_1(g))\circ \pi_1(f)\equiv rest_0^{f,x}\circ rest_0^{g,x} (y)
     \end{split}
\end{equation*}
and thus $func^{f,g}_0(y)$ is just reflexivity.

Using $func_n$, we now construct the proof $func^{f,g}_{n+1}(y) \colon (rest_{n+1}^{g\circ f,x}(y) = rest_{n+1}^{f,x}\circ rest_{n+1}^{g,x}(y))$.
By definition $rest_{n+1}^{g\circ f,x}(y)$ and $rest_{n+1}^{f,x}\circ rest_{n+1}^{g,x}(y)$, where $y:bnd^m_{n+1}(x)$,
compute to the following pairs:

\begin{equation*}
     \begin{split}
          rest_{n+1}^{g\circ f,x}\left(y,aug \right) \equiv %y:bnd^m_n(x),aug:\prod\limits_{h:\Delta(n+1,m)}Fill(rest_n^{h,x}(y))\right) =_{\beta} \\
           \left( rest_n^{g\circ f, x}(y), \lambda h:\Delta(n+1,k).subst^{Fill}_{func_n^{h,g\circ f}}(aug((g\circ f)\circ h))\right)
     \end{split}
\end{equation*}

\begin{equation*}
     \begin{split}
          rest_{n+1}^{f,x}\circ rest_{n+1}^{g,x}(y) \equiv rest_{n+1}^{f,x}\left(rest_n^{g,x}(y),
          \lambda h:\Delta(n+1,l).subst^{Fill}_{func_n^{h,g}}(aug(g\circ h))\right) \equiv \\
          \equiv \left(rest_n^{f,x}(rest_n^{g,x}(y)),\lambda h:\Delta(n+1,k).
          subst^{Fill}_{func_n^{h,f}}(subst^{Fill}_{func_n^{g,f\circ h}}(aug(g\circ(f\circ h))))\right)
     \end{split}
\end{equation*}
Each of these pairs is a point in a fiber of the family $\Pi F := (y:bnd^k_n(x))\prod\limits_{h:\Delta(n+1,k)}Fill(rest_n^{h,x}(y))$. By inductive
hypothesis, there is equality $func_n^{f,g}(y)$ between first components. Denote $aug_1$ and $aug_2$ the second components of the first and
the second pair respectively. It remains to prove, that $subst^{\Pi F}_{func_n^{f,g}}(aug_1)=aug_2$. By Lemma 2 and (LEqFE) rule this equality
is equivalent to $(h:\Delta(n+1,k))subst^{Fill}_{func_n^{f,g}}(aug_1\ h)=aug_2\ h$. $aug(g\circ(f\circ h))$ computationally equals to
$aug((g\circ f)\circ h)$ and by Lemma 1 $subst^{Fill}_{func_n^{h,f}}\circ subst^{Fill}_{func_n^{g,f\circ h}}=subst^{Fill}_{func_n^{h,f}\cdot func_n^{g,f\circ h}}$.
Thus $subst^{Fill}_{func_n^{f,g}}(aug_1\ h)$ and $aug_2\ h$ are two substitutions of the same term, therefore by (PI) they are equal.

\section{Implementation in Plastic}

Plastic is a light-weight proof assistant, developed by Paul Callaghan in 1999 \cite{plastic_page}. It's underlying system
is Luo's Logical Framework \cite{ZL:book94} and it's syntax is very similar to that of Lego \cite{legomanual:Report}. Plastic is the only proof
assistant that supports LTTs and is flexible enough for the developments, described in this paper.

In this section we outline the key aspects of the implementation of HoTTEq and SSTs in Plastic. The source code of Plastic
together with this implementation is available at github \cite{hott_plastic_page} (the SST implementation is located at the subdirectory 
\textbf{lib/Univalence/SimplicialTypes}).

\subsection{HoTTEq: Implementation in Plastic}

Rules of a type theory are expressed in Plastic by means of LF. For example, rules (LEqForm), (LEqIntro), (LEqElimT)
and (LEqElimP) for logical equality are
expressed as the following code in Plastic (here $A \rightarrow B$ is the abbreviation for the kind $(\_:A)B$):

\begin{lstlisting}
  [Eq : (A : Type) A -> A -> Prop];
  [Eqr : (A : Type) (a : A) Eq A a a];
  [EqE_T : (A : Type) (T : (x:A)(y:A)(_:Eq ? x y)Type)
           (_:(x:A)T x x (Eqr ? x))(x:A)(y:A)(p:Eq ? x y) T x y p];
  [EqE_P : (A : Type) (P : (x:A)(y:A)(_:Eq ? x y)Prop)
           (_:(x:A)P x x (Eqr ? x))(x:A)(y:A)(p:Eq ? x y) P x y p];
\end{lstlisting}

Plastic supports standard pattern of inductive definitions. For example, the type of paths is defined as follows:
\begin{lstlisting}
   Inductive [A:Type][Id_ : (x:El A)(y:El A)Type]
   Constructors
	[Idr : (z:El A)Id_ z z];
\end{lstlisting}

One of key features of Plastic is the mechanism to specify computation rules for customised eliminators of inductive types.
This makes it possible to implement induction of HoTTEq (and even higher inductive types). Specialised computation
rules can be defined by the command \textbf{SimpleElimRule}. This gives access to the underlying mechanism of Plastic, used for
inductive types and universes. Basically, the term, which is to be used as the combinator for elimination operator and
constructor arguments, is specifyed in this way. The general syntax is as follows:

\begin{lstlisting}
  SimpleElimRule TYPE_NAME ELIM_NAME ELIM_ARITY
           [CONSTR_1 CONSTR_1_ARITY = TERM : TYPE]
           ...
           [CONSTR_n CONSTR_n_ARITY = TERM : TYPE];
\end{lstlisting}

For example, this command allows one to define the recursion principle for a circle, so that it computes on the point constructor:
\begin{lstlisting}
 [Circle : Type];
 [base : Circle];
 [loop : Id_ ? base base];
 [CircRec : (A:Type)(a:A)(l:Id_ ? a a)Circle->A];

 SimpleElimRule Circle CircRec 4
        [base 0 = [A:Type][a:A][l:Id_ A a a] a :
                     (A:Type)(a:A)(_:Id_ A a a)A];
\end{lstlisting}

Because there's no way to implement the strong proof-irrelevance (PI) in the current version of Plastic, we replace it with
the following rule:
\begin{lstlisting}
 [PI : (P : Prop) (A : Type) (t : P -> A) (p , q : P)
                                      Eq ? (t p) (t q)];
\end{lstlisting}

\subsection{Semi-simplicial types}

The definition of SSTs by means of an inductive procedure, as described in \S\ref{sec:SST}, is implemented as two maps: $t:T,p:P(t)\vdash ih_T(t, p) \colon T$ and
$t:T,p:P(t)\vdash ih_P(t, p) \colon P(ih_T(t, p))$, where:

\begin{equation*}
  \begin{split}
    T := \sum\limits_{SST:\U_1}& \sum\limits_{bnd:SST\rightarrow \N\rightarrow \U_1}\prod\limits_{x:SST,k,m:\N, f:\Delta(k,m)}bnd^m(x)\rightarrow bnd^k(x)
  \end{split}
\end{equation*}
%and functoriality proof $sst\_rec_P$:

\begin{equation*}
  \begin{split}
  P((SST,bnd,rest):T) := \forall x:SST.\forall k, l, m:\N.\forall f:\Delta(k,l), g:\Delta(l,m). \\
  \forall y:bnd^m(x).rest^{g\circ f,x}(y) = rest^{f,x}(rest^{g,x}(y))
  \end{split}
\end{equation*}

Then, desired functions $sst\_rec_T \colon  \N \rightarrow T$ and $sst\_rec_P(n:\N) \colon P(sst\_rec_T(n))$, such that
$sst\_rec_T(n+1) \equiv ih_T(sst\_rec_T(n), sst\_rec_P(n))$, are obtained by applying induction principle of Section 3.
With proof scripts omitted as they are too cumbersome (see \cite{hott_plastic_page}) the code is as follows:

\begin{lstlisting}
 [T = Sigma ? ([SST:Type^1] Sigma ? ([bnd:(T^1 SST)==>(Nat==>Type^1)]
              Pi3 ? ? ? ([x:El (T^1 SST)][k:Nat][m:Nat]
              Pi ? ([f:El (Delta k m)]
              T^1 (ap2_ ? ? ? bnd x m) ==> T^1 (ap2_ ? ? ? bnd x k)
               )))) ];

 [P = [t:T] All_T ? ([x:(SST_pr t)]
            AllNat3 ([k,l,m:Nat]
            All_T ? ([f:El (Delta k l)] All_T ? ([g:El (Delta l m)]
            All_T ? ([y:(T^1 (bnd_pr t x m))]
            Eq ? (rest_pr t x k m (fm_comp k l m f g) y)
                 (rest_pr t x k l f (rest_pr t x l m g y)))))))];

 Claim ih_T : (n : Nat)(t : T)(p : (P t))T;
 ...

 Claim ih_P : (n : Nat)(t : T)(p : (P t))P (ih_T n t p);
 ...

 Claim t0 : T;
 ...

 Claim p0 : P t0;
 ...
\end{lstlisting}

And the type of all $n$-truncated SSTs is obtained by application of the data component \textbf{IE\_NatT} of induction for natural numbers:

\begin{lstlisting}
  pi1 ? ? (IE_NatT T P t0 p0 ih_T ih_P n);
\end{lstlisting}

\section{Conclusion}

The system HoTTEq and the succsessful definition of SSTs in HoTTEq, presented in this paper, is just one example of how 
HoTT can benefit from a logic enrichment. In general, strict \Prop{} of logic-enriched HoTT allows to talk about subsets 
$(\Gamma_{\Type},\Gamma_{\Prop})$ of the fibrant object, which represents a context of types $\Gamma_{\Type}$. In further work
we plan to explore in more detail a categorical semantics of the logic-enriched HoTT.

%The notion of logic in logic-enriched HoTT may be associated to
%strict \Prop, instead of hlogic. 

Definition of SSTs in HoTTEq clears up also many interesting directions for a further formalisation work.

Extending this definition to the definition of full simplicial types seems to us to be possible,
although it requires some technical details to be worked out.

Another possibility is to explore, what can be already done by means of SSTs. For example, straighforward definitins can be done for
 natural transformations between SSTs and complete semi-Segal types, which reflect the structure of weak $\infty$-category
without identities. In further work we plan to figure out, how the globular structure of the type of paths on a universe can
be realized as a complete semi-Segal type.

\begin{comment}
For example, complete semi-Segal types, which
reflect the structure of weak $\infty$-category without identities, may already be defined. A complete semi-Segal type
is a SST, such as:

\begin{itemize}
  \item Segal maps $X_n\rightarrow X_1\times_{X_0} \dots \times_{X_0} X_1$, assigning to a $n$-simplex the `composable' chain
    of 1-simplices in its boundary, is equivalence.
\end{itemize}
\end{comment}

\bibliography{main}

\end{document}